# AGNI: A radiative-convective model for lava planet atmospheres

**Harrison Nicholls** [1], **Raymond Pierrehumbert** [1], and **Tim Lichtenberg** [2]

**1** Department of Physics, University of Oxford, Parks Road, Oxford OX1 3PU, UK **2** Kapteyn Astronomical Institute, University of Groningen, P.O. Box 800, 9700 AV Groningen, The Netherlands

## Summary

It is important that we are able to accurately model the atmospheres of (exo)planets. This is because atmospheres play a central role in setting a planet's thermochemical environment at a given point in time, and also in regulating how it evolves over geological timescales. Additionally, it is primarily by observation of their atmospheres that we are able to characterise exoplanets. There is particular demand for accurate models in the context of so-called lava worlds: planets with molten interiors (or 'magma oceans').

AGNI is a Julia program designed to solve for the temperature and radiation environment within the atmospheres of rocky (exo)planets. It leverages a well established FORTRAN code (Edwards & Slingo, 1996; Sergeev et al., 2023) to calculate radiative fluxes from a given atmospheric temperature structure and composition, which – alongside representations of convection and other processes – enables an energy-conserving numerical solution for the atmospheric conditions. In contrast to most other numerical atmosphere models, AGNI uses a Newton-Raphson optimisation method to obtain its solution, which enables improved performance and scalability. Our model was specifically developed for use alongside planetary interior models within a coupled simulation framework. However, it can also be applied to scientific problems standalone when used as an executable program; it reads TOML configuration files and outputs figures and NetCDF datasets. AGNI can also function as a software library; it is used in this sense within the Jupyter notebook tutorials of our GitHub repository[1].

## Statement of need

It is thought that all rocky planets go through a 'magma ocean' stage, where their mantles are completely or largely molten (Elkins-Tanton & Seager, 2008; Lichtenberg & Miguel, 2024; Schaefer et al., 2016). For some planets this may be their permanent state, while for others it is fleeting. Magma oceans allow for rapid exchange of energy and volatiles between their atmospheres and interiors. Since this phase is likely common to many planets – including Earth and Venus – it is important that we understand the physical processes involved, and how these processes interact with each other (Maurice et al., 2024; Schaefer & Fegley, 2017). Accurate atmosphere models can allow us to connect the theory of these young planets to telescope observations, since it is primarily through their atmospheric properties that were are able to characterise them (Perryman, 2018; Piette et al., 2023). Modelling these young planets involves facing several poorly constrained quantites that govern their atmospheric composition (Guimond et al., 2023; Sossi et al., 2020). Recently, combined observations and modelling of exoplanet L 98-59 b have enabled the inference of a sulfur-rich atmosphere (Bello-Arufe et al., 2025). Hu et al. (2024) were able to characterise the atmosphere of 55 Cancri e by using an advanced but proprietary atmosphere model.

---

[1] nichollsh.github.io/AGNI

Several theoretical studies have modelled the atmospheres and evolution of these young planets, but all have made several simplifying assumptions. Lichtenberg et al. (2021) coupled a simple atmosphere climate model with an interior code to simulate magma ocean evolution, but they did not account for the possibility of convective stability in their atmospheres. Krissansen-Totton et al. (2021) and Krissansen-Totton et al. (2024) made similar assumptions. Selsis et al. (2023) used a pure-steam radiative-convective model to investigate convective stability, finding that it is likely to occur, but did not extend their work to coupled scenarios which explore the secular evolution of these planets or mixtures of gas species; it is unlikely that these atmospheres are exclusively composed of steam (Nicholls, Lichtenberg, et al., 2024). Piette et al. (2023) similarly explored potential atmospheres on observable lava planets, but did not consider the physics of atmosphere-interior coupling and chose semi-arbitrary gas compositions. Zilinskas et al. (2023) used the HELIOS code to model potential rock-vapour envelopes of sub-Neptune and super-Earth exoplanets, finding that the opacity of various gaseous species (notably SiO) plays a key role in determining their structure and observable properties. The demand for realistic modelling in the context of secular magma ocean evolution is apparent.

Ensuring sufficient spectral resolution is important in modelling the blanketing effect of these atmospheres, as resolving the opacity of their many gaseous components is known to be key in setting the rate at which these planets can cool (Nicholls, Lichtenberg, et al., 2024; Pierrehumbert, 2010). It is also important that we are able to run grids of models that explore the range of possible (and as-yet poorly constrained) conditions that these planets could exhibit, which demands efficient modelling given finite computational resources. Magma ocean crystallisation could take up to several Gyr in the presence of continuous tidal forcing and atmospheric blanketing (Driscoll & Barnes, 2015; Walterova & Behounkova, 2020). The numerical efficiency afforded by AGNI enables simulations of rocky planets over geological timescales as part of a coupled interior-atmosphere planetary evolution framework. AGNI has so far been used in Hammond et al. (2025), Nicholls, Pierrehumbert, et al. (2024), Nicholls, Guimond, et al. (2025), and Nicholls, Lichtenberg, et al. (2025).

## Comparison with other codes

AGNI is developed with the view of being coupled into the PROTEUS framework[2] alongside other modules. In addressing the aforementioned problems, it is able to:

- be self-coupled to a planetary interior model with an appropriate boundary condition,
- simulate atmospheres of diverse gaseous composition using realistic gas opacities and equations of state,
- solve for an atmospheric temperature structure that conserves energy and allows for convectively stable regions,
- operate with sufficient speed such that it may participate in the exploration of a wide parameter space.

These are possible due to the method by which AGNI numerically obtains a solution for atmospheric temperature structure and energy transport (Nicholls, Pierrehumbert, et al., 2024). Our model uses the Newton-Raphson method to conserve energy fluxes through each level of the column to a required tolerence. A typical runtime when applying the model standalone using its command-line interface (Figure 1b) with a poor initial guess of the true temperature profile is 3 minutes. When providing a 'good' guess, such as when AGNI is coupled within the PROTEUS framework (Figure 1a), an atmosphere solution will be obtaind in less than 1 minute. A single radiative transfer calculation takes approximately 30 ms, performed under the correlated-k and two-stream approximations using SOCRATES[3]: a well-established FORTRAN code developed by the UK Met Office (Amundsen et al., 2017; Edwards & Slingo, 1996;

---

[2] fwl-proteus.readthedocs.io
[3] github.com/nichollsh/SOCRATES

Sergeev et al., 2023). Convection, condensation, and sensible heat transport are also accounted for.

HELIOS[4] (Malik et al., 2017) is a popular atmosphere model similar to AGNI, but it depends on an Nvidia GPU in order to perform radiative transfer calculations. Whilst this makes each calculation fast, it also means that HELIOS cannot be used on platforms without an Nvidia GPU or with limited resources. GENESIS (Piette et al., 2023) has been applied to lava planet atmospheres but is closed-source and not publically available. Exo_k (Selsis et al., 2023) is open source and written in pure Python, but not designed to be coupled with an interior evolution model. These codes have been used to model the atmospheres of static non-evolving planets, while AGNI stands out as being the only open source code currently integrated into a comprehensive interior-atmosphere evolution framework like PROTEUS. No other models of lava planet atmospheres implement a real-gas equation of state.

Coupling with PROTEUS is one primary use-case for AGNI. However, our model can also be used standalone (as in Hammond et al. (2025)) through its command-line interface and configuration files, or through Jupyter notebooks (as in the tutorials). Figure 1 below compares these two primary use-cases driving the development of AGNI.

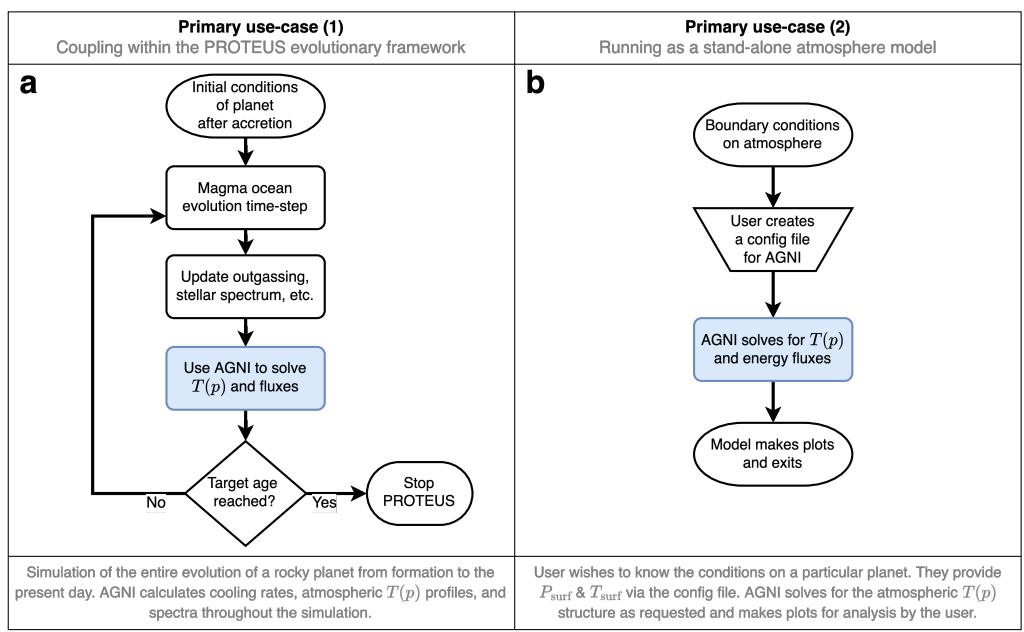

Figure 1: A visual comparison of the two primary use-cases for AGNI. Left: coupling within the PROTEUS framework. Right: using the code standalone.

## Similar tools

One of the key strengths of AGNI is that it is designed to be used as part of the PROTEUS framework, which can simulate the evolution of rocky planets on Gyr timescales (Figure 1a). Other radiative-convective atmosphere models – compared to AGNI in the previous section – are available. HELIOS, for example, has been used alongside volatile and rock-vapour outgassing models (e.g. LavAtmos[5], van Buchem et al. (2025)) to simulate the emission spectra of lava planets Seidler, Fabian L. et al. (2024).

---

[4] github.com/exoclime/HELIOS
[5] github.com/cvbuchem/LavAtmos

## Future developments

The landscape of exoplanet science is rapidly evolving. Future updates to AGNI may include:

- Incorporation of aerosols and hazes; supported by SOCRATES in principle, but currently not configurable through AGNI
- An equatorial multi-column mode which parametrises zonal redistribution by atmospheric dynamics
- Accounting for compositional inhibition of convection; e.g. via the Ledoux stability criterion
- Parametrisation of dry convection with a 'full spectrum' model which better represents turbulence in convective fluids
- Parallel computing, allowing multiple solvers to run simultaneously and to enhance the speed of each radiative transfer calculation.

## Documentation

The documentation for AGNI can be read online at [nichollsh.github.io/AGNI](https://nichollsh.github.io/AGNI).